\documentclass[12pt]{article}

\usepackage{amssymb}
\usepackage{amsmath}
\usepackage{amsbsy}
\usepackage{amscd}
\usepackage{amsfonts}
\usepackage{amsthm}
\usepackage{mathrsfs}
\usepackage{verbatim}
\usepackage{hyperref}
\usepackage{fullpage}
\usepackage{mathdots}
\usepackage{graphicx,subfigure}
\usepackage[english]{babel}
\usepackage[utf8]{inputenc}
\usepackage{tikz}
\usepackage{adjustbox}
\usepackage{authblk}
\usepackage{caption}
\usepackage{multicol}
\usepackage{mathtools}
\usepackage{mathtext}
\usepackage{stackengine,scalerel}

\usepackage{tikz-cd}
\usetikzlibrary{backgrounds,fit, matrix}
\usetikzlibrary{positioning}
\usetikzlibrary{calc,through,chains}
\usetikzlibrary{arrows,shapes,snakes,automata, petri}
\usepackage[boxsize=1.25em, centerboxes]{ytableau}

\usetikzlibrary{trees}

\newtheorem{Theorem}[equation]{Theorem}
\newtheorem{Corollary}[equation]{Corollary}
\newtheorem{Lemma}[equation]{Lemma}
\newtheorem{Proposition}[equation]{Proposition}

\theoremstyle{definition}
\newtheorem{Definition}[equation]{Definition}
\newtheorem{Example}[equation]{Example}

\newtheorem{Remark}[equation]{Remark}

\numberwithin{equation}{section}
\numberwithin{figure}{section}

\newcommand{\F}{{\mathbb F}}

\newcommand{\Z}{{\mathbb Z}}

\newcommand{\N}{{\mathbb N}}

\newcommand{\mc}[1]{\mathcal{#1}}

\newcommand{\mbf}[1]{\mathbf{#1}}

\usepackage[all,2cell]{xy}
	\usepackage{amssymb, hyperref}
	\usepackage[all,2cell]{xy}
	\usepackage{hyperref}
	\usepackage{array}
	\usepackage[capitalise]{cleveref}

\DeclareMathOperator{\Mat}{\mathbf{Mat}}

\usepackage{graphicx}
\begin{document}
	
\title{Generalized Hyperderivative Reed-Solomon Codes}

\author[1]{Mahir Bilen Can}
\author[2]{Benjamin Horowitz}

\affil[1]{\small{Department of Mathematics, Tulane University, USA\\mahirbilencan@gmail.com}}
\affil[2]{\small{Department of Mathematics, Tulane University, USA\\bhorowitz1@tulane.edu}}
\maketitle

\maketitle

\begin{abstract}
\medskip

This article introduces Generalized Hyperderivative Reed-Solomon codes (GHRS codes), which generalize NRT Reed-Solomon codes. 
Its main results are as follows: 1) every GHRS code is MDS, 2) the dual of a GHRS code is also an GHRS code, 3) determine subfamilies of GHRS codes whose members are low-density parity-check codes (LDPCs), and 4) determine a family of GHRS codes whose members are quasi-cyclic. 
We point out that there are GHRS codes having all of these properties.

\medskip

\medskip
		

\end{abstract}

\section{Introduction}
\label{S:MDS}

In modern coding theory, low density parity-check (LDPC) codes have revolutionized error correction by combining excellent performance with efficient decoding algorithms. At the same time, classical algebraic constructions, such as Generalized Reed-Solomon codes, are celebrated for their maximum distance separable (MDS) property but inherently lack the sparsity needed for LDPC implementations. In this paper, we introduce a novel family of codes, dubbed \emph{Generalized Hyperderivative Reed-Solomon codes} (GHRS codes), that strikingly reconcile these two desirable features. Moreover, our construction not only generalizes the NRT metric Reed-Solomon codes but also yields Quasi-Cyclic codes whose parity-check matrices exhibit the low-density structure crucial for practical applications. We believe that this unexpected fusion of MDS optimality, LDPC sparsity, quasi-cyclicity, as well being closed under duality operations, opens up exciting new avenues for both theoretical exploration and real-world coding applications for the family of GHRS codes. We now proceed to detail the construction and present main results of our work.

We will use the following notation throughout the paper. Let $\mathbb{Z}_+$ denote the set of positive integers, and let $\mathbb{F}_q$ denote the finite field with $q$ elements. Positive integers $r$, $s$, and $t$ will satisfy $r \leq q$ (assuming $q$ is fixed) and $t \leq rs$. For $t \in \mathbb{Z}_+$, let $\mathcal{P}(t-1)$ represent the space of polynomials $f(z) \in \mathbb{F}_q[z]$ with degree at most $t-1$. The $\mathbb{F}_q$-vector space of all $s \times r$ matrices with entries in $\mathbb{F}_q$ will be denoted by $\Mat_{s \times r}(\mathbb{F}_q)$.  

Our construction is based on an $r$-tuple $\alpha := (\alpha_1, \dots, \alpha_r) \in \mathbb{F}_q^r$, referred to as the list of \emph{evaluation points}, and an $s \times r$ matrix $V :=(v_{ij})_{i=1,\dots,s \atop j=1,\dots,r}$ with entries in $\mathbb{F}_q$, called the \emph{multiplier matrix}.
Using these ingredients, we introduce the \emph{Generalized Hyperderivative Reed-Solomon (GHRS) code}, denoted $GHRS(\alpha,V,t-1)$, as the image of the following evaluation map:
\begin{align*}
Ev_{ \alpha, V } : \mathcal{P}(t-1) &\longrightarrow \Mat_{s\times r}(\F_q) \\
f(z) &\longmapsto 
\begin{bmatrix}
v_{1,1} \partial^{ 0 } f(\alpha_1) & v_{1,2} \partial^{ 0 } f(\alpha_2) & \cdots & v_{1,r} \partial^{ 0 } f(\alpha_r) \\
v_{2,1} \partial^{ 1 } f(\alpha_1) & v_{2,2} \partial^{ 1 } f(\alpha_2) & \cdots & v_{2,r} \partial^{ 1 } f(\alpha_r) \\
\vdots & \vdots & \ddots & \vdots \\
v_{s, 1} \partial^{ s-1 } f(\alpha_1) &v_{s, 2} \partial^{ s-1 } f(\alpha_2) & \cdots & v_{s, r} \partial^{ s-1 } f(\alpha_r) \\
\end{bmatrix}.
\end{align*}
Here, $\partial^{ i } f$ stands for the $i$-th hyperderivative of $f$, which we define precisely in the preliminaries section.
\medskip

When the multiplier matrix $V$ is the all-ones matrix, GHRS codes reduce to the {\it classical NRT Reed-Solomon codes}, first introduced by Rosenbloom and Tsfasman in their seminal paper~\cite{RT1997}. Although Rosenbloom and Tsfasman did not explicitly use the term ``NRT Reed-Solomon codes,'' they described these codes as analogs of Reed-Solomon codes in the $m$-metric. This $m$-metric, now widely referred to as the NRT metric, owes its name to Niederreiter's foundational work~\cite{Niederreiter1987,Niederreiter1991} (see, for instance, \cite{FirerAlvesPinheiroPanek}).  
For consistency with contemporary conventions, we will adopt the term ``NRT metric'' throughout this work. 
A formal definition will be provided in the preliminaries section.

We refer to the analogs of the Reed-Solomon codes in the NRT metric as {\em NRT Reed-Solomon codes}.
After~\cite{RT1997}, they were investigated by Skriganov~\cite{Skriganov2001} for uniform distributions, by Niederreiter and Xing~\cite{NX}
for digital nets. 
Further generalization were found by Niederreiter and \"Ozbudak in~\cite{NiOz2002}, and more recently in~\cite{CanMonteroOzbudak}. 
Related Reed-Solomon type constructions involving multiplicities and ordered metrics,
though developed from a different perspective, were also investigated by Zhou, Lin,
and Abdel-Ghaffar in \cite{ZLA}.

\medskip

To kick off the presentation of the main results of our paper, we begin with extending the well-known fact that classical NRT Reed-Solomon codes are maximum distance separable (MDS) codes in the NRT metric. This extension motivated our exploration of properties of GHRS codes that remain hidden in the specialized case of classical NRT Reed-Solomon codes.

Let $\mathcal{C} \subseteq \mathbf{Mat}_{s\times r}(\mathbb{F}_q)$ be a linear code endowed with the NRT metric $d_{C(s,r)}$.
The subscript $C(s,r)$ in the distance notation represents the poset consisting of $r$ disjoint chains of height $s$.
This choice of notation will be explained in the preliminaries section. 
Let $d_{C(s,r)}(\mathcal{C})$ denote the minimum distance of $\mathcal{C}$ with respect to $\mathcal{C}$. 
We know from~\cite[Theorem 1]{RT1997} that there is a Singleton-type bound in the NRT metric for $\mathcal{C}$: 
\begin{equation}\label{eq:singleton}
\dim(\mathcal{C}) + d_{C(s,r)}(\mathcal{C}) \leq n + 1.
\end{equation}
We say that $\mathcal{C}$ is a \emph{maximum distance separable (MDS) code with respect to $d_{C(s,r)}$} if the inequality in~(\refeq{eq:singleton}) is an equality:
\[
\dim(\mathcal{C}) + d_{C(s,r)}(\mathcal{C}) = n + 1.
\]

\begin{Theorem}\label{intro:T1}[MDS Property]
Let $\alpha=(\alpha_1,\dots, \alpha_r)$ be a list of $r$ distinct evaluation points from $\F_q$. 
Let $V= (v_{ij})_{i=1,\dots,s \atop j=1,\dots,r}$ be a multiplier matrix. 
If every entry of $V$ is nonzero, then $GHRS(\alpha,V,t-1)$ is a $t$-dimensional MDS code with respect to the NRT metric. 

In particular, the minimum distance of $GHRS(\alpha,V,t-1)$ with respect to the NRT metric is given by 
$$
d_{C(s,r)}(GHRS(\alpha,V,t-1)) = rs - t+1.
$$
\end{Theorem}

\medskip

\medskip

We know from the work~\cite{HyunKim} of Hyun and Kim that, for poset metric codes 
duality must be formulated with respect to dual posets. 
Since the underlying NRT poset $C(s,r)$ is self-dual as a poset, in the present article, passing to the dual poset does not change
the combinatorial structure. Nevertheless, the associated metric is
complementary, and this convention is implicitly respected in all duality
results stated in the paper.
Thus, our dual codes are defined using the standard dot product on
$\mathbf{Mat}_{s\times r}(\mathbb{F}_q)$,
\[
 A\cdot B  := \sum_{i=1}^s\sum_{j=1}^r a_{ij}b_{ij},
\]
which is equivalent, via vectorization, to the trace inner product on $\mathbb{F}_q^{sr}$.
In parts of the poset/ordered-metric literature, one sometimes uses a bilinear form tailored to the
underlying translation association scheme (rather than the coordinatewise dot/trace inner product); see,
for instance \cite{DoughertySkriganov2002}. In the present paper we keep the standard dot product
(and hence the equivalent trace inner product under vectorization) because it yields the usual linear
dual code and is exactly the notion of orthogonality used in our duality statement in Section~\ref{S:Duality}.

Our next result extends the classical duality theorem for {\em Generalized Reed-Solomon codes}, which are defined as follows:
$$
GRS(\alpha,V,t-1):= \{ ( v_1 f(\alpha_1),\dots, v_r f(\alpha_r))\in \F_q^r \mid f(z) \in \mc{P}(t-1) \},
$$
where $\alpha=(\alpha_1,\dots, \alpha_r)$ is a list of distinct evaluation points from $\F_q$, and $V=\begin{bmatrix} v_1 & \cdots & v_r\end{bmatrix}$ is a multiplier matrix from $\Mat_{1\times r}(\F_q)$.
For a detailed introduction to the Generalized Reed-Solomon codes and their history we recommend~\cite[Section 5]{Roth}.

To state our main duality result, we maintain the notation of our previous theorem. 
Then the dimension of the code $GHRS(\alpha,V,rs-2)$ is $rs-1$. 
The dual code of this code is defined by 
$$
GHRS(\alpha,V,rs-2)^\perp := \{ A\in \Mat_{s\times r}(\F_q) \mid A\cdot B = 0 \text{ for all $B\in GHRS(\alpha,V,rs-2)$} \}.
$$
Since $\dim GHRS(\alpha,V,rs-2)^\perp = 1$, there exists a matrix $W\in \Mat_{s\times r}(\F_q)$ such that every codeword of $GHRS(\alpha,V,rs-2)^\perp$ is a scalar multiple of $W$. 
In particular, we have 
\begin{align}\label{A:givesthedual}
GHRS(\alpha,V,rs-2)^\perp = GHRS(\alpha,W,0).
\end{align}
In this notation, our next result is the following.

\begin{Theorem}
\label{T:Hij-isabasis}\label{intro:T2}[Main Duality Theorem]
Let $\alpha=(\alpha_1,\dots, \alpha_r)$ be a list of $r$ distinct evaluation points from $\F_q$. 
Let $V= (v_{ij})_{i=1,\dots,s \atop j=1,\dots,r}$ be a multiplier matrix. 
Let $W$ be the multiplier matrix that gives the dual of the code $GHRS(\alpha,V,rs-2)$ as in (\ref{A:givesthedual}).
Then the dual of the code $GHRS(\alpha,V,t-1)$ with respect to NRT metric is the Generalized Hyperderivative Reed-Solomon code $GHRS(\alpha,W,rs-t)$:
$$
GHRS(\alpha,V,t-1)^{\perp} = GHRS(\alpha,W,rs-t-1).
$$
Indeed, setting $t=rs-1$ gives $GHRS(\alpha,V,rs-2)^{\perp}=GHRS(\alpha,W,0)$, which agrees with~(\ref{A:givesthedual}).

\end{Theorem}

\bigskip

The proof of our Theorem~\ref{intro:T2} is intriguing in its own right. 
The classical Hermite interpolation basis consists of functions that construct an interpolating polynomial matching both the values and the derivatives of a function at specified points. In the proof of Theorem~\ref{intro:T2}, we employ a similar, yet more general construction. 
More precisely, we develop a `Generalized Hermite Interpolation Polynomial Basis,'  
$$
\{H'_{ij} (z) \in \mc{P}(rs-1) \mid i=0,\dots, s-1,\ j=1,\dots, r\}, 
$$
which serves as an analogue to the basis of $\Mat_{s\times r}(\F_q)$ formed by elementary matrices.
The basis vectors are determined relative to the evaluation map $Ev_{\alpha,V}$. 
It is worth mentioning here that Hermite interpolation polynomials was previously considered by Skriganov in~\cite[Section 5]{Skriganov2001} in the more specific context of NRT Reed-Solomon codes. 
In fact, his work focused solely on the existence of solutions to the Hermite interpolation problem. 
In this regard, while Skriganov's results laid important groundwork for the NRT Reed-Solomon codes, our work extends these foundations by explicitly constructing bases that are specifically tailored to our framework and applications. 
For further details, we refer the reader to the proof of Theorem~\ref{intro:T2}.
\bigskip

Next, we will discuss the parity-check matrices of Hyperderivative Reed-Solomon codes. 
A code $\mc{C} \subseteq \mathbb{F}_q^N$ is said to be a \emph{low-density parity-check (LDPC) code} if it can be defined as the kernel of a sparse parity-check matrix.
Here, by a {\em sparse matrix}, we mean a matrix whose number of nonzero entries is less than its number of zero entries. 
\medskip

It is straightforward to show that, with respect to the Hamming metric, an MDS code cannot be an LDPC code since every set of $n-k$ columns of any parity-check matrix must be linearly independent, forcing more than half of the matrix entries to be nonzero (see, for example, \cite[Proposition 3.4]{BlaumRoth} for a detailed proof). 
We use our previous result to show that we can always find LDPC codes among the Generalized Hyperderivative Reed-Solomon codes.
More specifically, we obtained the following result.

\begin{Theorem}\label{intro:T3}[LDPC Property]
Let $r,s$, and $t$ be integers such that $r,s,t\geq 2$ and $t \leq rs-1$. 
Let $\alpha=(\alpha_1,\dots, \alpha_r)$ be a list of $r$ distinct evaluation points from $\F_q$. 
Let $V= (v_{ij})_{ i=1,\dots,s\atop j=1,\dots, r}$ be a multiplier matrix such that 
$v_{i,j}\neq 0$ for every $i\in \{1,\dots, s\}$ and $j\in \{1,\dots, r\}$. 
Then the following assertions hold:
\begin{enumerate}
\item[(1)] If $t=s$ and $r+1 \leq s$, then the code $GHRS(\alpha,V,rs-t)$ is LDPC.
\item[(2)] If the inequality $st \geq  t^2 + t + r+1$ holds, then the code $GHRS(\alpha,V,rs-t)$ is LDPC.
\end{enumerate}
\end{Theorem}

\bigskip

A linear code $\mathcal{C} \subseteq \mathbb{F}_q^N$ is said to be \emph{quasi-cyclic} of index $\ell$ if for any codeword $c = (c_0, c_1, \ldots, c_{N-1}) \in \mathcal{C}$, the $\ell$-cyclic shift 
\[
T^\ell_{\F_q^N} (c):= (c_{N-\ell}, c_{N-\ell+1}, \ldots, c_{N-1}, c_0, c_1, \ldots, c_{N-\ell-1})
\]
is also in $\mathcal{C}$. 
Quasi-cyclic codes are widely used in digital communications and data storage due to their excellent error-correcting capabilities and efficient encoding and decoding algorithms. They are also important in network coding and quantum computing for optimizing data transmission and protecting quantum information.
It turns out that, appropriately reorganized, our GHRS codes are quasi-cyclic codes of index $r$. 
This is our next result.

\begin{Theorem}\label{intro:T4}[Quasi-cyclic Property]
Let $\alpha \in\mathbb{F}_q^*$ be such that $\{1, \alpha, \alpha^2, \dots, \alpha^{r-1}\}$ is a cyclic subgroup of $\F_q^*$. 
Let $u:=(1, \alpha,\alpha^2, \dots,\alpha^{r-1})$.
Let $V:=(v_{ij})_{i=1,\dots, s\atop j=1,\dots, r}$ be an $s\times r$ multiplier matrix such that 
\[
\frac{v_{ij}}{v_{i,j-1}} = \alpha^{\, i-1} \quad \text{ for all $i=1,\dots,s$ and $j=1,\dots,r$},
\]
with the convention that $v_{i,0} = v_{i,r}$. 
Then $GHRS(u,V,t-1)$ is a quasi-cyclic code of index $r$. 
\end{Theorem}

\medskip

\begin{Remark}
To justify the final sentence of the abstract of this paper, we fix three integers $r,s$, and $t$ such that $r,s,t\geq 2$ and $t \leq rs-1$. 
Let $\alpha$ be a nonzero element from a sufficiently large finite field.
We assume that $\{1, \alpha, \alpha^2, \dots, \alpha^{r-1}\}$ is a cyclic subgroup of $\F_q^*$. 
Let $u:=(1, \alpha, \alpha^2, \dots, \alpha^{r-1})$.
Let $V:=(v_{ij})_{i=1,\dots, s\atop j=1,\dots, r}$ be an $s\times r$ multiplier matrix such that 
\[
\frac{v_{ij}}{v_{i,j-1}} = \alpha^{\, i-1} \quad \text{ for all $i=1,\dots,s$ and $j=1,\dots,r$},
\]
with the convention that $v_{i,0} = v_{i,r}$.
If, in addition, either of the following conditions hold: 
\begin{itemize}
\item $t=s$ and $r+1 \leq s$, or 
\item $st \geq  t^2 + t + r+1$,
\end{itemize}
then the Generalized Hyperderivative Reed-Solomon code $GHRS(u,V,t-1)$ has the following properties:
\begin{enumerate}
\item MDS with respect to the NRT metric; 
\item LDPC; 
\item quasi-cyclic.
\end{enumerate}
\end{Remark}
\bigskip

The structure of our paper is as follows. 
In the next (Preliminaries) section, we introduce our basic objects, including poset metrics and discuss NRT codes.
In Section~\ref{S:MDS}, we prove that the GHRS codes are MDS (Theorem~\ref{intro:T1}).
In Section~\ref{S:Duality} we prove the duality property of the family of GHRS codes (Theorem~\ref{intro:T2}). 
In Section~\ref{S:LDPC} we determine some families of LDPC GHRS codes (Theorem~\ref{intro:T3}).
In Section~\ref{S:Quasicyclic}, we prove that some GHRS codes are quasi-cyclic (Theorem~\ref{intro:T4}).

\section{Preliminaries and Notation}
\label{S:Preliminaries}

We begin our preliminaries section with a discussion of hyperderivatives.
To facilitate this, it is useful to clearly define the binomial coefficients. 
For $(n, a) \in \mathbb{Z} \times \mathbb{Z}$, the binomial coefficient $\binom{n}{a}$ is defined as follows:
\[
\binom{n}{a} = 
\begin{cases}
\frac{n (n-1) \cdots (n-a+1)}{a!} & \text{if } a > 0, \\
1 & \text{if } a = 0, \\
0 & \text{if } a < 0.
\end{cases}
\]
We note in passing that although binomial coefficients are well-known, it is important to define their limits explicitly, as certain recursive arguments involving binomial coefficients with different extremal conventions might lead to conflicting conclusions (see, for example, \cite{Sorensen, CanJoshuaRavindraII}, as well as \cite{GhorpadeLudhani} for clarification).

\subsection{Hyperderivatives.}

Let $t$ be a positive integer. 
Let $f(x)\in \mathcal{P}(t-1)$ be given in the form $f(x) = f_0 + f_1 x + \cdots + f_{t-1} x^{t-1}$. 
Then the {\em $j$-th hyperderivative} of $f(x)$ is the polynomial defined by
\begin{align}\label{A:hyperderivative}
\partial^j f(x) := {0\choose j} f_0 x^{-j} + {1\choose j} f_1 x^{1-j} + \cdots + {t-1\choose j} f_{t-1} x^{t-1-j}.
\end{align}
It is not difficult to see that $\partial^j f(x)$ is the coefficient of $z^j$ in $f(x+z)$, that is, 
\begin{equation}\label{eq:Hassederivative}
\partial^j f(x)\;=\;[z^j]\; f(x+z).
\end{equation}

\begin{Example}\label{E:directly}
Let $f(x)= (x-u)^t$ for some $u\in \F_q$ and $t\in \Z_+$.
Let $a\in \N$. Then we have 
\begin{align}\label{A:productrule}
\partial^a f (x) =
\begin{cases} 
{t\choose a} (x-u)^{t-a} & \text{ if $0\leq a \leq t$},\\
0 & \text{ if $a > t$}.
\end{cases}
\end{align}
This formula follows directly from the expansion of ${t\choose n} (x-u)^{t-n}$ and (\ref{A:hyperderivative}). 
In particular, we have 
\begin{align}\label{A:productrule1}
\partial^a f (u):= \partial^a f (x)\Big\vert_{x=u} =
\begin{cases} 
1 & \text{ if $a=t$},\\
0 & \text{ if $a\neq t$}.
\end{cases}
\end{align}
\end{Example}

\bigskip

One of the most important properties of hyperderivatives is that they allow Taylor series expansion over finite fields.
More precisely, let $f(x)$ be a polynomial with coefficients from $\F_q$. 
Let $u\in \F_q$. 
We denote by $\nu_f(u)$ the order of vanishing of $f(x)$ at $u$. 
This means that the highest power of $(x-u)$ that divides $f(x)$ is $\nu_f(u)$.
In other words we have 
$$
f(x) = (x-u)^{\nu_f(u)}g(x)
$$ 
for some polynomial $g(x)$ such that $g(u)\neq 0$. 
\medskip

Then the {\em Taylor expansion of $f(x)$ at $u$} is given by 
\begin{align*}
f(x) = \sum_{j=0}^{t-1} \partial^j f(u) (x-u)^j.
\end{align*}
We mention two direct consequences of this expansion:
\begin{enumerate}
\item[(1)] For all $m\in \N$, we have 
\begin{align}\label{A:nu=0iff}
\nu_f(u) = m \iff \text{$\partial^j f(u) = 0$ for all $j\in \{0,1,\dots, m-1\}$ but $\partial^m f(u) \neq 0$. }
\end{align}
\item[(2)] If the degree of $f(x)$ is less than the order of the hyperderivative we are taking, then we get zero: 
\begin{align}\label{A:exceeds}
\partial^a f(x) = 0\quad \text{ if $a > \deg (f(x))$}.
\end{align}
\end{enumerate}

For other interesting properties of the hyperderivatives, we refer readers to the useful monograph~\cite{NiederreiterLidl}.

\subsection{$s$-jets and the evaluation map.}

Let $f\in \F_q[x]$. 
The \emph{$s$-jet} of $f$ at $u\in\F_q$ is the column vector of Hasse derivatives
\[
J^{s-1}_u(f) \;:=\; \big(\partial^0 f(u),\,\partial^1 f(u),\,\dots,\,\partial^{s-1} f(u)\big)^\top \in \F_q^{\,s}.
\]
Equivalently, $J^{s-1}_u(f)$ records the class of $f$ in $\F_q[x]/(x-u)^s$ with respect to the basis $1,(x-u),\dots,(x-u)^{s-1}.$

We now introduce our notion of a \emph{multiplier matrix} as a matrix 
$V=(v_{i,j})\in \Mat_{s\times r}(\F_q)$ that scales the jet entries. 
We write $v_{\bullet,j}=(v_{1,j},\dots,v_{s,j})^\top$ for the $j$-th column of $V$.

The GHRS-codes build on the following evaluation map. 
Given distinct evaluation points $u:=(\alpha_1,\dots,\alpha_r)$, the evaluation map is
\[
Ev_{\alpha,V}:\ \mc P(t\!-\!1)\longrightarrow \Mat_{s\times r}(\F_q),\qquad
f \longmapsto A=(A_{i,j}),\ \ A_{i,j}=v_{i,j}\,\partial^{\,i-1}f(\alpha_j).
\]
Thus the $j$-th column of $A$ is the scaled $s$-jet 
$$
v_{\bullet,j}J^{s-1}_{\alpha_j}(f) := \big( v_{1,j} \partial^0 f(u), \cdots , v_{s,j} \partial^{s-1} f(u) \big)^\top.
$$
We refer to such $s$-dimensional columns as \emph{jet blocks}.

\subsection{NRT metrics.}

Classical error-correcting codes (ECC) utilize the Hamming metric.
For our codes, we use the NRT-metrics, which are defined as follows. 
\medskip

Let $(P,\leq)$ be a poset whose underlying set is given by $[n]:=\{1,\dots, n\}$. 
The {\em $P$-metric on $\F_q^n$}, denoted $d_P$, is defined by the assignment 
\begin{align*}
(v,w) \longmapsto \omega_P(v-w)
\end{align*}
for $v,w\in \F_q^n$. 
Here, $\omega_P(v-w)$, called the \emph{$P$-weight of $v-w$}, is the number of $j\in [n]$ such that $j\leq i$ for some $i \in {\rm supp}(v-w) := \{ i\in \{1,\dots, n\} \mid v_i \neq w_i \}$.

We proceed to explain how NRT metric arise as an example of a poset metric. 
Let $\mathbf{a}:=(a_1,\dots, a_s)\in \F_q^s$.
We convert $\mathbf{a}$ into a column matrix by taking its reverse-transpose: 
$$
A:=
\begin{bmatrix}
a_s \\
a_{s-1}\\
\vdots \\
a_1
\end{bmatrix}.
$$
In this notation, the weight of $\mathbf{a}$ is given by $s- i+1$, where $i$ is index of the first nonzero row of $A$. 
For example, we have 
$$
\omega \left(
\begin{bmatrix}
0 \\ 0 \\ 0 \\ 1 \\ 0 \\ 1\\ 1\\ 0
\end{bmatrix}
\right) = 8 - 4 +1 = 5.
$$

The poset metric interpretation of this weight is obtained as follows.
Let $r,s\in \Z_+$ be such that $n = rs$.
Recall that $C(s,r)$ denotes the union of $r$ disjoint chains, each containing $s$ vertices. 
We label the vertices of $C(s,r)$ from 1 to $n$ as follows:
\begin{enumerate}
\item Label the smallest elements of each chain by using the numbers $1,\dots, r$ starting from the left most chain towards right. 
\item Move to the next level and repeat.
\end{enumerate}
For example, in Figure~\ref{F:C(2,3)}, we depict $C(2,3)$ with vertices labeled just as defined. 
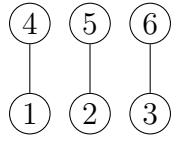
\begin{figure}[htp]
\begin{center}
\begin{tikzpicture}[scale=.8, every node/.style={circle, draw, fill=white, inner sep=2pt}]
    \node (1) at (0,0) {1};
    \node (2) at (0,1.5) {4};
    \node (3) at (1,0) {2};
    \node (4) at (1,1.5) {5};
    \node (5) at (2,0) {3};
    \node (6) at (2,1.5) {6};
    \draw (1) -- (2);
    \draw (3) -- (4);
    \draw (5) -- (6);
\end{tikzpicture}
\end{center}
\caption{The Hasse diagram of $C(2,3)$.}
\label{F:C(2,3)}
\end{figure}

By using our previous discussion, we now define the NRT-metric on $\Mat_{s\times r}(\F_q)$. 
Initially, let $A$ be a column vector with $s$ entries. 
As before, let $i$ denote the index of the first nonzero entry of $A$. 
Then we set, 
\begin{align*}
\omega_{C(s,1)}(A) :=  s-i+1,
\end{align*}
Next, let us assume that $A$ is an $s\times r$ matrix, $A:=(a_{ij})_{i=1,\dots,s \atop j=1,\dots, r}\in \Mat_{s\times r}(\F_q)$.
Let $A_1,\dots, A_r$ denote the columns of $A$. 
Then the {\em NRT-weight} of the matrix $A$, denoted $\omega_{C(s,r)}(A)$, is defined as 
$$
\omega_{C(s,r)}(A):= \omega_{C(s,1)}(A_1)+\cdots + \omega_{C(s,1)}(A_r). 
$$
It is easy to check that this sum is the value of the $P$-weight of $A$ where $P$ is the poset $C(s,r)$. 
Hereafter, we denote the corresponding NRT-metric by $d_{C(s,r)}$. 
We call a code $\mc{C}$ together with NRT-metric on it, an {\em NRT-code}.

\medskip

\begin{Example}
For $s=1$, a code in $\Mat_{s\times r}(\F_q)$ is simply a code in $\F_q^r$, and the NRT-metric is the same as the Hamming metric.
Furthermore, in this case, the Generalized Hyperderivative Reed-Solomon code specializes to the Generalized Reed-Solomon code.
In particular, the NRT-weight of a codeword $(f(\alpha_1),\dots, f(\alpha_r))$, where $f\in \mc{P}(t-1)$, is simply the number of nonzero coordinates.
\end{Example}

Next, we present the formula for the NRT-weight of a codeword $Ev_{\alpha,V}(f) \in \Mat_{s\times r}(\F_q)$, for arbitrary $s$.

\medskip

\begin{Lemma}\label{L:weightoff}
Let $A$ denote the matrix $Ev_{\alpha,V}(f)$, where the multiplier matrix $V$ has no zero entries. 
If $A_j$ denotes the $j$-th column of $A$, then the NRT-weight of $A_j$ is given by 
$$
\omega_{C(s,1)}(A_j) = s-\nu_f(\alpha_j).
$$
Therefore, the NRT-weight of the matrix $A$ is given by 
\begin{align}\label{A:weightoff}
\omega_{C(s,r)}(Ev_{\alpha,V}(f)) = \sum_{j=1}^r (s-\nu_f(\alpha_j)) = rs  - \sum_{j=1}^r \nu_f(\alpha_j).
\end{align}
\end{Lemma}

\section{GHRS Codes are MDS}
\label{S:MDS}

The main result of this section is Theorem~\ref{intro:T1} which extends the MDS property of the classical Reed-Solomon codes. 
We recall the statement for convenience of the reader: 
\bigskip

Let $\alpha=(\alpha_1,\dots, \alpha_r)$ be a list of $r$ distinct evaluation points from $\F_q$. 
Let $V= (v_{ij})_{i=1,\dots,s \atop j=1,\dots,r}$ be a multiplier matrix. 
If every entry of $V$ is nonzero, then $GHRS(\alpha,V,t-1)$ is a $t$-dimensional MDS code with respect to the NRT metric. 

In particular, the minimum distance of $GHRS(\alpha,V,t-1)$ with respect to the NRT metric, denoted $d_{C(s,r)}(GHRS(\alpha,V,t-1))$, is given by 
$$
d_{C(s,r)}(GHRS(\alpha,V,t-1)) = rs - t+1.
$$
\medskip

\begin{proof}[Proof of Theorem~\ref{intro:T1}]
Let $f$ be a polynomial of degree at most $t-1$ such that $Ev_{\alpha,V}(f)$ is the zero of $\Mat_{s\times r}(\F_q)$. 
Since every entry of $V$ is nonzero, it follows that $\partial^i f(\alpha_j) = 0$ for all $i\in \{0,\dots, s-1\}$ and $j\in \{1,\dots, r\}$. 
As will be shown later (Theorem~\ref{T:Hij'isabasis}), the interpolation basis yields a unique
expansion, implying injectivity of the evaluation map.
Hence, we have $f(x) = 0$, proving the injectivity of the evaluation map $Ev_{\alpha,V}$.
In particular, now we know that 
$$
\dim GHRS(\alpha,V,t-1) = \dim \mc{P}(t-1) = t.
$$
We proceed to show that the minimum NRT weight of our code is $rs - t+1$.

\bigskip

Let $f(x)$ be a polynomial of degree at most $t-1$. 
By Lemma~\ref{L:weightoff}, the NRT weight of $Ev_{\alpha,V}(f)$ is given by 
$\omega_{C(s,r)}(Ev_{\alpha,V}(f)) = \sum_{j=1}^r (s-\nu_f(\alpha_j)) = rs  - \sum_{j=1}^r \nu_f(\alpha_j)$.
Notice that this expression of the weight of $Ev_{\alpha,V}(f)$ is independent of the multiplier matrix $V$.
Hence, the NRT weights of the matrices 
$Ev_{\alpha,V}(f)$ and $Ev_{\alpha,M}(f)$, where $M=(1)_{i=1,\dots,s \atop j=1,\dots,r}$ is the all-ones matrix, are the same. 
It follows that the minimum weight of the `classical NRT Reed-Solomon code' $GHRS(\alpha,M,t-1)$ is the same as the minimum distance of the 
Generalized Hyperderivative Reed-Solomon code $GHRS(\alpha,V,t-1)$.
But we already know that $GHRS(\alpha,M,t-1)$ is an MDS code, hence, 
$d_{C(s,r)}(GHRS(\alpha,M,t-1)) = rs-t+1$. 
Therefore, we obtain that $$d_{C(s,r)}(GHRS(\alpha,M,t-1)) = rs-t+1,$$ finishing the proof of our theorem. 
\end{proof}

\section{Interpolation Basis and Duality}
\label{S:inter}

Let $r,s\in \Z_+$. 
In this section we introduce a special interpolation basis for the space of polynomials $\mc{P}(rs-1)$. 
To this end, we fix $r$ distinct evaluation points $\alpha_1,\dots, \alpha_r$ from $\F_q$.
For every $i \in \{0,\dots, s-1\}$ and $j\in \{1,\dots, r\}$, we define 
\begin{align}\label{A:PandL}
P_{i,j}(x):= (x-\alpha_j)^i \quad\text{ and }\quad L_{j}(x):=  \prod_{\substack{l=1,\dots,r\\ l\neq j}} \frac{ (x-\alpha_l)^{s}}{(\alpha_j-\alpha_l)^{s}}.
\end{align}
\medskip
Evidently, we have 
$$
L_j(\alpha_m)=\begin{cases} 1 &\text{ if $j=m$},\\ 0 & \text{otherwise.}\end{cases}
$$
It is also evident that 
\begin{align*}
0\leq \deg (P_{i,j}(x)) = i\leq s-1 \quad\ \text{ and }\ \quad \deg (L_j) = s(r-1).
\end{align*}
Finally, for each $i\in\{0,\dots,s-1\}$ and $j\in\{1,\dots,r\}$, we define the \emph{interpolation basis} element,  
\[
H_{i,j}(x) := P_{i,j}(x)\,L_j(x) = (x-\alpha_j)^i\,L_j(x).
\]

\begin{Lemma}
We maintain our previous notation.
Let $s,r\in \Z_+$. 
Then $\{ H_{i,j} (x) \}_{i=0,1,\dots, s-1\atop j=1,\dots, r}$ is a basis for $\mc{P}(rs-1)$. 
\end{Lemma}
\begin{proof}
Since there are $rs$ polynomials in the set $\{ H_{i,j} (x) \}_{i=0,1,\dots, s-1\atop j=1,\dots, r}$, 
it suffices to show that it is a linearly independent set. 
We prove this by using mathematical induction on $s$.

If $s=1$, then $H_{i,j}(x) = L_j(x)$. 
But $L_j(x)$'s are Lagrange interpolation polynomials and their linear independency is well-known. 
We assume that our assertion holds for $s-1$, and we proceed to show it for $s$. 

Next, assume that we have a linear relation of the form 
$$
\sum_{i=0}^{s-1} \sum_{j=1}^r c_{i,j} H_{i,j}(x) = 0.
$$
We separate the sum as follows: 
$$
\sum_{i=0}^{s-1} \sum_{j=1}^r c_{i,j} H_{i,j}(x) = \underbrace{ \left( \sum_{i=0}^{s-2} \sum_{j=1}^r c_{i,j} H_{i,j}(x) \right)}_{A(x)} + \underbrace{\left( \sum_{j=1}^r c_{s-1,j}H_{s-1,j}(x)\right)}_{B(x)}.
$$
Each polynomial $H_{s-1,j}(x)$ in $B(x)$ vanishes at $x = \alpha_j$ with order $s-1$ for $j = 1, \dots, r$, while the polynomials in $A(x)$ vanish at $x = \alpha_j$ with orders $i \in \{0, 1, \dots, s-2\}$. 
Thus, $A(x) + B(x) = 0$ if and only if $A(x) = 0$ and $B(x) = 0$.
Now we can apply the induction hypothesis to $A(x)$ to conclude that $c_{i,j} = 0$ for all $i \in \{0, 1, \dots, s-2\}$ and $j = 1, \dots, r$. 
Next, we show that $c_{s-1,j} = 0$ for all $j = 1, \dots, r$. 
To this end, note that $H_{s-1,j}(x)$ has an order of vanishing $s-1$ at $\alpha_j$, while it vanishes to order $s$ at $x = \alpha_l$ for $l \neq j$. 
Consequently, the equation 
\[
c_{s-1,1} H_{s-1,1}(x) + \cdots + c_{s-1,r} H_{s-1,r}(x) = 0
\]
holds if and only if $c_{s-1,1} = \cdots = c_{s-1,r} = 0$. 
This completes the proof of the lemma.
\end{proof}

\begin{Proposition}\label{P:Hintroduced}
We maintain the notation from the previous lemma. 
In addition, we assume that $s>1$ and $r>1$.
Let $i \in \{0,\dots, s-1\}$ and $j\in \{1,\dots, r\}$.
Then we have 
\[
\partial^k H_{i,j}(\alpha_j)=
\begin{cases}
0, & 0\le k<i,\\
L_j(\alpha_j), & k=i,\\
\partial^{k-i}L_j(\alpha_j), & i<k\le i+s(r-1),\\
0, & k>i+s(r-1).
\end{cases}
\]
\end{Proposition}

\begin{proof}
Recall that the hyperderivatives evaluated at a point $u$ give the coefficients of the Taylor series expansion of a polynomial at the point $u$. 
Let $\alpha_{k,j}$ denote $\partial^k L_j(\alpha_j)$. 
Then we have 
\[
L_j(x)=1+\alpha_{1,j}(x-\alpha_j)+\alpha_{2,j}(x-\alpha_j)^2+\cdots + \alpha_{(r-1)s,j}(x-\alpha_j)^{(r-1)s}.
\]
It follows that 
\[
H_{i,j}(x)=(x-\alpha_j)^iL_j(x)=(x-\alpha_j)^i+\alpha_{1,j}(x-\alpha_j)^{i+1}+\alpha_{2,j}(x-\alpha_j)^{i+2}+\cdots+\alpha_{(r-1)s,j}(x-\alpha_j)^{i+(r-1)s}.
\]
The rest of the proof follows from the definition of the hyperderivative defined in (\ref{eq:Hassederivative}).
\end{proof}

\bigskip

\begin{Corollary}\label{C:HasseofHij}
Let $s,r\in \Z_+$ be such that $r> 1$, $s>1$. 
Let $k,i\in \{0,\dots, s-1\}$ and $j,l\in \{1,\dots, r\}$.
If $k\leq i $, then we have 
\[
\partial^k H_{i,j}(\alpha_l) =\delta_{k,i} \delta_{j,l}.
\]
\end{Corollary}

\begin{proof}
Let $l\in \{1,\dots, r\} \setminus \{j\}$.
Since $H_{i,j}(\alpha_l) = 0$, the Taylor series expansion of $H_{i,j}(x)$ at $x= \alpha_l$ starts at $(x-\alpha_l)^s$.
This means that 
\[
\partial^k H_{i,j}(\alpha_l) =0\quad \text{for all } k=0,\dots,s-1,
\]
Next, we assume that $l=j$. 
Then we know from Proposition~\ref{P:Hintroduced}) that 
\[
\partial^k H_{i,j}(\alpha_j)=\delta_{i,k}\quad \text{for all } k=0,\dots,i.
\]
Hence, the proof follows. 
\end{proof}

\bigskip

We wish to expand polynomials $f(x)$ in a basis where the coefficients are given by the values of the hyperderivatives. 
Unfortunately, Proposition~\ref{P:Hintroduced} indicates that 
\[
\partial^k H_{i,j}(\alpha_j)\neq \delta_{i,k}\quad \text{for all } k=i+1,\dots,r(s-1),
\]
since the coefficients $\alpha_{k-i,j}$ (for $k>i$) of the Taylor expansion of $L_j(x)$ may not be 0 or 1.
Our next result shows the existence of an appropriate basis that has the desired property.

\begin{Theorem}\label{T:Hij'isabasis}
We maintain our previous notation. 
Then there exists a basis 
\[
\{ H_{i,j}' (x) \in \mc{P}(rs-1) \mid i = 0,\dots,s-1, \ j=1,\dots, r\}
\]
such that $\partial^kH'_{i,j}(\alpha_j)=\delta_{k,i}$ for $i,k=0,\dots,s-1$.
In particular, every polynomial $f(x)$ of degree at most $rs-1$ has a unique expansion of the form 
\[
f(x)=\sum_{j=1}^r\sum_{i=0}^{s-1} \partial^if(\alpha_j)\,H'_{i,j}(x).
\]
\end{Theorem}
\medskip

{\em We caution the reader about the indices used in the proof.
For an $s\times r$ matrix, the entries are customarily indexed by $(i,j)$,
where $1\leq i \leq s$ and $1\leq j \leq r$. However,
in our matrices, the $(i,j)$-th entry corresponds to the $(i-1)$-th hyperderivative
of a polynomial evaluated at the $j$-th evaluation point. 
Therefore, depending on the convenience of notation, $i$ varies from $0$ to $s-1$.}

\begin{proof}
For each $j\in \{1,\dots, r\}$, we define an $s\times s$ matrix $M$ whose $(i,k)$-th entry is given by 
\[
M_{i,k} :=\partial^{i-1} H_{k-1,j}(\alpha_j),\quad i,k=1,\dots,s.
\]
It is easy to check from Proposition~\ref{P:Hintroduced} that $M$ is lower triangular with ones on the diagonal and hence invertible.
Now, for each $j$, we introduce a new basis functions $\{H'_{r,j}(x)\}_{r=0}^{s-1}$ by setting 
\[
\begin{bmatrix}
H'_{0,j}(x)\\[1mm]
H'_{1,j}(x)\\[1mm]
\vdots\\[1mm]
H'_{s-1,j}(x)
\end{bmatrix}
:=
M^{-1}\,
\begin{bmatrix}
H_{0,j}(x)\\[1mm]
H_{1,j}(x)\\[1mm]
\vdots\\[1mm]
H_{s-1,j}(x)
\end{bmatrix}.
\]
Then we have 
\begin{align}\label{A:beforesubs}
H'_{k-1,j}(x)=\sum_{t=1}^{s} (M^{-1})_{k,t}\,H_{t-1,j}(x),\quad k=1,\dots,s.
\end{align}

For $i=1,\dots, s$, we apply the hyperderivative of order $i-1$ and evaluate at $x= \alpha_j$:
\[
\partial^{i-1} H'_{k-1,j}(\alpha_j) = \sum_{t=1}^{s} (M^{-1})_{k,t}\,\partial^{i-1}H_{t-1,j}(\alpha_j).
\]
Since $\partial^{t-1} H_{i-1,j}(\alpha_j)=M_{t,i}$, it follows that
\[
\partial^{i-1} H'_{k-1,j}(\alpha_j)=\sum_{k=1}^{s} (M^{-1})_{k,t}\, M_{t,i}.
\]
But the sum on the right is exactly the $(k,i)$-th entry of the identity matrix, $M^{-1}M$, which equals $\delta_{k,i}$. 
This finishes the proof of our first assertion. 
\medskip

For our second assertion, we expand $f(x) \in \mc{P}(rs-1)$ in our new basis: 
\begin{align}\label{A:mainexpansionoff}
f(x)=\sum_{j=1}^r\sum_{i=0}^{s-1} c_{i,j} H'_{i,j}(x).
\end{align}
Since $\{H_{i,j}'(x) \mid i=0,\dots, s-1,\ j=1,\dots, r\}$ is a basis, the coefficients $c_{k,j}$, $k=0,\dots, s-1$, $j=1,\dots, r$, are uniquely determined. 
Applying the $k$-th hyperderivative to (\ref{A:mainexpansionoff}), and using the fact that $\partial^k H'_{i,j}(\alpha_j)=\delta_{i,k}$, 
we see that $\partial^i f(\alpha_j) = c_{i,j}$. 
This completes the proof of our theorem. 
\end{proof}

We have an analog of Corollary~\ref{C:HasseofHij} with a more relaxed hypothesis.

\begin{Corollary}\label{C:HasseofH'}
Let $i,k\in \{1,\dots, s\}$ and $j,l \in \{1,\dots, r\}$. 
Then we have 
\[
\partial^{i-1} H'_{k-1,j}(\alpha_l)=\delta_{i,k}\delta_{j,l}.
\]
\end{Corollary}

\begin{proof}
We begin with the assumption that $l=j$ so that $\delta_{l,j}=1$. 
Then by Theorem~\ref{T:Hij'isabasis} we have
$$
\partial^{i-1}H'_{k-1,l}(\alpha_j)=\delta_{k,i}\quad \text{for } i,k=1,\dots,s.
$$
We proceed with the assumption that $l \in \{1,\dots, r\}\setminus \{j\}$.
Hence, we have $\delta_{l,j} =0$. 
In this case, we apply the hyperderivative of $(i-1)$-th order to (\ref{A:beforesubs}) and then evaluate the result at $x=\alpha_j$: 
$$
\partial^{i-1} H'_{k-1,j}(\alpha_l)=\sum_{t=1}^{s} (M^{-1})_{k,t}\,\partial^{i-1} H_{t-1,j}(\alpha_l),\quad k=1,\dots,s.
$$
Then, as we showed in the (first sentence of the) proof of Corollary~\ref{C:HasseofHij}, the term $\partial^{i-1} H_{t-1,j}(\alpha_l)$ is 0 for all $i=1,\dots, s$. 
In other words, we have $\partial^{i-1} H'_{k-1,j}(\alpha_l)=0$ for $l\neq j$. 
It follows that $\partial^i H'_{k,j}(\alpha_l)= \delta_{i,k}\delta_{j,l}$. 
\end{proof}

\bigskip

\begin{Definition}
The {\em standard basis} for the space of $s\times r$ matrices is the basis 
$$
\{E_{i,j} \mid 1\leq i \leq s,\ 1\leq j \leq r\},
$$
where $E_{i,j}$ is the $s\times r$ elementary matrix with 1 at the $(i,j)$-th position and 0 elsewhere. 
\end{Definition}

We close this section by recording an important consequence of our previous corollary.

\medskip

\begin{Proposition}\label{P:H'isabasis}
For $r,s\in \Z_+$, 
let $\alpha:=\{\alpha_1,\dots, \alpha_r\}$ be a set of distinct evaluation points from $\F_q$, and $V:=(v_{i,j})_{\substack{i=0,\dots, s-1,\\ j=1,\dots, r}}$ a multiplier matrix.
Then the images under the evaluation map $Ev_{\alpha,V} : \mc{P}(rs-1)\to \mbf{Mat}_{s\times r}(\F_q)$ of the polynomials 
$$
(1/{v_{i,j}})H'_{i,j}(x), 
$$ 
for $i=0,\dots, s-1,\ j=1,\dots, r$, give the standard basis for the space of matrices $\mathbf{Mat}_{s\times r}(\F_q)$. 
\end{Proposition}

\begin{proof}
The {\em $(k,l)$-th elementary matrix}, denoted $E_{k,l}$, is the matrix that has 1 at its $(k,l)$-th position and 0's elsewhere. 
Clearly, $\{E_{k,l}\}_{\substack{k=1,\dots, s\\ l=1,\dots, r}}$ is a basis for $\mbf{Mat}_{s\times r}(\F_q)$.
We notice that 
$$
Ev_{\alpha,V} \left((1/{v_{i,j}}) H'_{i,j}(x)\right) = \Bigl( \partial^k H'_{i,j} (\alpha_l)\Bigr)_{\substack{k=0,\dots, s-1\\ l=1,\dots, r}}.
$$
But our previous corollary shows that these matrices are precisely the elementary matrices in $\mbf{Mat}_{s\times r}(\F_q)$.
Since we have the ``correct number'', that is, $rs$ many of them, we see that 
$Ev_{\alpha,V} \left((1/{v_{i,j}}) H'_{i,j}(x)\right)$ form a basis.
Hence, our proof follows.
\end{proof}

\bigskip

\subsection{Duality}
\label{S:Duality}

In this subsection we determine the dual of a GHRS code.
We show that the highest-degree coefficient in the product $f(x)g(x)$ (when $f \in P(t-1)$ and $g \in P(rs-t)$) can be expressed in terms of the Hasse derivatives of $f$ and $g$, and that this coefficient is zero. This is the critical step in proving that the trace-inner product of the evaluations (with appropriate multipliers) vanishes, hence establishing the duality.
Here, the trace-inner product on the space of $ s \times r $ matrices over $\mathbb{F}_q$ is defined by
\[
\langle A, B \rangle = \operatorname{Tr}(A^\top B),
\]
for $ A, B \in \mathbf{Mat}_{s\times r}(\mathbb{F}_q) $, where $ A^\top $ is the transpose of $ A $ and $ \operatorname{Tr} $ denotes the matrix trace. 
This coincides with the dot product defined in the Introduction under the standard
identification of matrices with vectors:
\[
\langle A, B \rangle = A \cdot B = \sum_{i=1}^s \sum_{j=1}^r A_{i,j} B_{i,j}.
\]

\bigskip

We now consider a degree $rs-1$ GHRS code $\mc{C}:=GHRS(\alpha,V,rs-2)$,
where $\alpha:=\{\alpha_1,\dots, \alpha_r\}$ is a set of distinct evaluation points from $\F_q$ and $V=(v_{i,j})_{i=1,\dots, s\atop j=1,\dots, r}$ is the corresponding multiplier matrix. 
Let $\mc{C}^\perp$ denote the dual of $\mc{C}$ in $\Mat_{s\times r}(\F_q)$ with respect to trace-inner product.
Since $\dim \mc{C} = rs-1$ and since the ambient matrix space is $rs$ dimensional, we know that $\dim \mc{C}^\perp =1$. 
Therefore, there exists a matrix $W=(w_{i,j}) \in \Mat_{s\times r}(\F_q)$ such that 
every element of $\mc{C}^\perp$ is a scalar multiple of $W$. 
Also, by the duality between $\mc{C}$ and $\mc{C}^\perp$, for every polynomial $h(x)$ of degree at most $rs-2$, we have 
\begin{align}\label{A:EvVf1}
Ev_{\alpha,V}(h)\cdot W =0.
\end{align}
The left hand side of (\ref{A:EvVf1}) is given by 
\begin{align}\label{A:EvVf2}
\sum_{i=1}^{s} \sum_{j=1}^r \partial^{i-1} h(\alpha_j) v_{i,j} w_{i,j}.
\end{align}

We are now ready to prove our second main result mentioned in the introduction, that is, the {\em Duality Theorem.}
We recall its statement for the convenience of readers. 
\bigskip

Let $\alpha=(\alpha_1,\dots, \alpha_r)$ be a list of $r$ distinct evaluation points from $\F_q$. 
Let $V= (v_{ij})_{i=1,\dots,s \atop j=1,\dots,r}$ be a multiplier matrix. 
Let $W$ be the multiplier matrix that gives the dual of the code $GHRS(\alpha,V,rs-2)$ as in (\ref{A:givesthedual}).
Then the dual of the code $GHRS(\alpha,V,t-1)$ is the Generalized Hyperderivative Reed-Solomon code $GHRS(\alpha,W,rs-t)$:
$$
GHRS(\alpha,V,t-1)^{\perp} = GHRS(\alpha,W,rs-t-1).
$$

\bigskip

\begin{proof}[Proof of Theorem~\ref{intro:T2}]
Let $f(x)$ (resp. $g(x)$) be a polynomial of degree at most $t-1$ (resp. $rs-t$).
We will show that the trace-inner product of $Ev_{\alpha,V}(f) $ and $Ev_{\alpha,W}(g)$ is zero:
\begin{align}\label{A:wantzero}
Ev_{\alpha,V}(f) \cdot Ev_{\alpha,W}(g) &= \sum_i \sum_j v_{i,j} w_{i,j} \partial^{i-1} f (\alpha_j) \partial^{i-1} g (\alpha_j) = 0
\end{align}

We fix an $s\times r$ multiplier matrix $U=(u_{i,j})$ with coordinates all 1: 
$$
u_{i,j}=1 \quad \text{for } i=1,\dots,s,\ j=1,\dots, r.
$$
Proposition~\ref{P:H'isabasis} shows that the images of the basis vectors 
$H'_{i,j}(x)$ for $\mc{P}(rs-1)$ under $Ev_{\alpha,U}$ are the elementary matrices $E_{i+1,j}$, providing us with the standard basis for $\Mat_{s\times r}(\F_q)$. 
In particular, for any data $\gamma_{i,j} \in \F_q$, where $i=1,\dots, s$ and $j=1,\dots, r$,
there exists a unique polynomial $h(x)$ of degree at most $rs-1$ such that 
\begin{align}\label{A:hexists}
Ev_{\alpha,U}(h(x)) = (\gamma_{i,j})_{i=1,\dots, s \atop j=1,\dots, r}.
\end{align}
We now fix our data as follows:
\begin{align*}
\gamma_{i,j} :=  \partial^{i-1} f (\alpha_j) \partial^{i-1} g (\alpha_j), 
\end{align*}
where $f(x)$ and $g(x)$ are as in the previous paragraph. 
Let $h(x)$ be the unique polynomial for which (\ref{A:hexists}) holds for this data.
Then, by the definition of our evaluation map $Ev_{v,U}$, we have 
\begin{align}\label{A:actuallyzero}
\partial^{i} h(\alpha_j) =  \partial^{i} f (\alpha_j) \partial^{i} g (\alpha_j)
\end{align}
for all $i=0,\dots, s-1$ and $j=1,\dots, r$. 
Since $\deg h(x) \leq rs-1$, we see from (\ref{A:EvVf1}) and (\ref{A:EvVf2}) that,
for our choice of $h(x)$, we have
\begin{align*}
Ev_{\alpha,V}(f) \cdot Ev_{\alpha,W}(g) &= \sum_i \sum_j v_{i,j} w_{i,j} \partial^{i-1} f (\alpha_j) \partial^{i-1} g (\alpha_j) \\
&= \sum_i \sum_j v_{i,j} w_{i,j} \partial^{i-1} h(\alpha_j) && \text{(by (\ref{A:EvVf2}))}\\
&=0 && \text{(by (\ref{A:EvVf1}))}.
\end{align*}
This finishes the proof of our theorem.
\end{proof}

\section{Parity-Check Matrices}\label{S:LDPC}

Low density parity check (LDPC) codes are traditionally defined using \emph{Tanner graphs}, which are bipartite graphs representing the structure of a parity-check matrix. A \emph{Tanner graph} \( G = (V, C, E) \) associated with the parity check matrix $H=(H_{i,j})$ consists of:
\begin{itemize}
    \item A set \( V \) of \emph{variable nodes}, each corresponding to a coordinate of the codeword.
    \item A set \( C \) of \emph{check nodes}, each corresponding to a parity-check equation.
    \item An edge \( (v_i, c_j) \in E \) exists if the \( i \)-th variable participates in the \( j \)-th parity-check equation. In other words, there is an edge between the nodes corresponding to $v_i$ and $c_j$ if the entry \( H_{j,i} \) of the parity-check matrix \( H \) is nonzero.
\end{itemize}

A code is called \emph{LDPC} if its Tanner graph is \emph{sparse}, meaning that both the variable node degrees and check node degrees are bounded by constants independent of the blocklength.
More precisely, we say that a linear code over $\F_q$ is \emph{LDPC} if it admits a parity-check matrix $H$ whose associated Tanner graph has bounded left- and right-degrees; each column has weight at most $w_c = O(1)$ and each row has weight at most $w_r = O(1)$ independent of the blocklength.
Here, the notation $O(1)$ means that the quantity is bounded by an absolute constant that does not depend on the blocklength $n$. In other words, the column and row weights remain uniformly bounded as $n \to \infty$.
(All degrees below are with respect to the natural block structure of $s$-jets.)

\subsection*{Decoding note (belief propagation over jet blocks)}
On the Tanner graph induced by $H$, standard sum-product (or min-sum) decoding can be applied at the level of \emph{jet blocks}. Messages are $s$-dimensional beliefs over the jet variables at each evaluation point $\alpha_j$. When $s$ is small (the regime of interest here), each check-node update costs $O(s^2)$ operations via precomputed convolution tables for the linear constraints coming from the hyperderivative relations, so one decoding iteration costs $O(r\,s^2)$. A straightforward pseudocode can be implemented along these lines; we omit the listing for brevity and focus on the structural bounds and complexity.

In this section, we will discuss the sparsity of the parity-check matrices of our GHRS codes.
Although there is a description of the general form of a generator matrix for an NRT code in~\cite{Alves2011}, we will take a more direct approach.
\medskip

As before, we fix a multiplier matrix $V=(v_{i,j})_{\substack{1\leq i \leq s,\\ 1\leq j \leq r}}$ and a list of evaluation points 
$\alpha:=(\alpha_1,\dots, \alpha_r)$.
Let $\mc{B}:=\{f_1(x),\dots, f_t(x)\}$ be a basis for $\mc{P}(t-1)$.
Then a generator matrix for $GHRS(\alpha,V,t-1)$ is a $t\times rs$ block matrix $F$ of the form 
\begin{align}
F(\mc{B}):= \begin{bmatrix}
Ev_{\alpha,V}(f_1(x)) \\
Ev_{\alpha,V}(f_2(x)) \\
\vdots  \\
Ev_{\alpha,V}(f_t(x))
\end{bmatrix} \in \Mat_{t\times rs}(\F_q). 
\end{align}
In particular, for the canonical basis $\mc{B}_{on}:=\{1,x,\dots, x^{t-1}\}$ of $\mc{P}(t-1)$, we have 
\begin{align}\label{A:eachblock}
Ev_{\alpha,V}(x^m) =
\begin{bmatrix}
v_{1,1} \alpha_1^m & v_{1,2} \alpha_2^m & \cdots & v_{1,r} \alpha_r^m \\
v_{2,1} {m \choose 1} \alpha_1^{m-1} & v_{2,2} {m \choose 1} \alpha_2^{m-1} & \cdots & v_{2,r} {m \choose 1} \alpha_r^{m-1} \\
\vdots & \vdots & \ddots & \vdots \\
v_{s-2, 1} {m \choose s-2} \alpha_1^{m-(s-2)} & v_{s-2, 2}  {m \choose s-2} \alpha_2^{m-(s-2)} & \cdots & v_{s-2, r} {m \choose s-2} \alpha_r^{m-(s-2)}\\
v_{s-1, 1} {m \choose s-1} \alpha_1^{m-(s-1)} & v_{s-1, 2}  {m \choose s-1} \alpha_2^{m-(s-1)} & \cdots & v_{s-1, r} {m \choose s-1} \alpha_r^{m-(s-1)}
\end{bmatrix}
\end{align}
for $m=0,\dots, t-1$.
We identify the block $Ev_{\alpha,V}(x^m)$ of $F(\mc{B}_{on})$ by a row vector, denoted $F_{m+1}$, by expanding $Ev_{\alpha,V}(x^m)$ in the elementary matrix basis of $\Mat_{t\times rs}(\F_q)$.
In other words, $F_{m+1}$ is the row vector that is obtained from $Ev_{\alpha,V}(x^m)$ by reading its entries row-by-row from left-to-right, top-to-bottom starting at the first row. 
Hence, under these identifications, the generator matrix $F(\mc{B}_{on})$ looks as follows: 
\begin{align}\label{A:mainblock}
\resizebox{\textwidth}{!}{$
\left[
\begin{array}{c|c|c|c|c}
v_{1,1}\ \cdots\ v_{1,r} & 0\ \cdots\ 0 & 0\ \cdots\ 0 & \cdots & 0 \\
v_{1,1}\alpha_1\ \cdots\ v_{1,r}\alpha_r & v_{2,1}\binom{1}{1}\ \cdots\ v_{2,r}\binom{1}{1} & 0\ \cdots\ 0 & \cdots & 0 \\
v_{1,1}\alpha_1^2\ \cdots\ v_{1,r}\alpha_r^2 & v_{2,1}\binom{2}{1}\alpha_1\ \cdots\ v_{2,r}\binom{2}{1}\alpha_r & v_{3,1}\binom{2}{2}\ \cdots\ v_{3,r}\binom{2}{2} & \cdots & 0 \\
\vdots & \vdots & \vdots & \ddots & \vdots \\
v_{1,1}\alpha_1^{t-1}\ \cdots\ v_{1,r}\alpha_r^{t-1} & v_{2,1}\binom{t-1}{1}\alpha_1^{t-2}\ \cdots\ v_{2,r}\binom{t-1}{1}\alpha_r^{t-2} & v_{3,1}\binom{t-1}{2}\alpha_1^{t-3}\ \cdots\ v_{3,r}\binom{t-1}{2}\alpha_r^{t-3} & \cdots & v_{s,r}\binom{t-1}{s-1}\alpha_r^{t-s}
\end{array}
\right]
$}
\end{align}
It will be useful to view (\ref{A:mainblock}) as a block matrix $\begin{bmatrix} M_0 & \cdots & M_{s-1} \end{bmatrix}$, where each $M_i$ ($i \in\{ 0, \dots, s-1\}$) is a $t \times r$ matrix. 
\bigskip

When a basis for a code is given, applying invertible elementary row operations to it results in another generator matrix of the code. We inductively apply invertible elementary row operations to the matrix $F(\mathcal{B}_{on})$, starting with the first row and working towards the lower rows. This process transforms $\begin{bmatrix} M_0 & \cdots & M_{s-1} \end{bmatrix}$ into a block matrix $\begin{bmatrix} \widetilde{M}_0 & \cdots & \widetilde{M}_{s-1} \end{bmatrix}$, which is in row echelon form. 
Furthermore, for each $i \in \{1, \dots, s-1\}$, similar to the block $M_i$, the first $i$ rows of the new block $\widetilde{M}_i$ are zero.

\medskip

We proceed to determine a lower bound for the number of 0's in the matrix 
$\begin{bmatrix} \widetilde{M}_0 & \cdots & \widetilde{M}_{s-1} \end{bmatrix}$. 
\medskip

\textbf{Case 1.} We assume that $2\leq s \leq t$. 
\medskip

Since $\widetilde{M}_0$ is in row echelon form, the number of 0's it contains 
can be calculated by subtracting from $rt$ the number of entries in the $r\times r$ upper triangular part: 
\begin{align}\label{A:zerosofM_0}
rt - \frac{(r+1)r}{2}.
\end{align}
Now let $i\in \{1,\dots, s-1\}$. 
Since the first $i$ rows of $\widetilde{M}_i$ are zero, the number of zeros in $\widetilde{M}_i$ is bounded from below by the number 
\begin{align}\label{A:otherzerosinM_i}
i r.
\end{align}
Combining (\ref{A:zerosofM_0}) and (\ref{A:otherzerosinM_i}), we find that 
a lower bound for the number of 0's in the row echelon matrix $\begin{bmatrix} \widetilde{M}_0 & \cdots & \widetilde{M}_{s-1} \end{bmatrix}$ is given by 
\begin{align}\label{A:lowerboundCase1}
rt - \frac{(r+1)r}{2} + r + 2r +\cdots + (s-1)r = rt - \frac{(r+1)r}{2}+ r \frac{s(s-1)}{2}.
\end{align}
\medskip

\textbf{Case 2.} We assume that $1 < t<s$. 
\medskip

In this case, the last $s-t$ blocks in $\begin{bmatrix} \widetilde{M}_0 & \cdots & \widetilde{M}_{s-1} \end{bmatrix}$ are 0 matrices. 
These blocks account for 
\begin{align}\label{A:Case2a}
rt(s-t)
\end{align}
zeros. 
Additionally, the block matrix $\begin{bmatrix} \widetilde{M}_0 & \cdots & \widetilde{M}_{t-1} \end{bmatrix}$, can be treated as in Case 1 with $t=s$. 
Hence, it contributes at least 
\begin{align}\label{A:Case2b}
 rt - \frac{(r+1)r}{2}+ r \frac{t(t-1)}{2} 
\end{align}
zeros. 
Therefore, combining (\ref{A:Case2a}) and (\ref{A:Case2b}), we find that the number of zeros is at least 
\begin{align}\label{A:Case2}
rt(s-t)+  rt - \frac{(r+1)r}{2}+ r \frac{t(t-1)}{2} =
rst - \frac{rt^2}{2} - \frac{rt}{2} - \frac{(r+1)r}{2}.
\end{align}

\bigskip

We are now ready to prove our third main result, Theorem~\ref{intro:T3}, from the introduction. 
We recall its statement for convenience. 
\medskip

Let $r,s$, and $t$ be integers such that $r,s,t\geq 2$ and $t \leq rs-1$. 
Let $\alpha=(\alpha_1,\dots, \alpha_r)$ be a list of $r$ distinct evaluation points from $\F_q$. 
Let $V= (v_{ij})_{i=1,\dots,s \atop j=1,\dots,r}$ be a multiplier matrix such that 
$v_{i,j}\neq 0$ for every $i\in \{1,\dots, s\}$ and $j\in \{1,\dots, r\}$. 
Then the following assertions hold:
\begin{enumerate}
\item[(1)] If $t=s$ and $r+1 \leq s$, then the code $GHRS(\alpha,V,rs-t)$ is LDPC.
\item[(2)] If the inequality $st \geq  t^2 + t + r+1$ holds, then the code $GHRS(\alpha,V,rs-t)$ is LDPC.
\end{enumerate}

\medskip

\begin{proof}[Proof of Theorem~\ref{intro:T3}]

First, we will prove (1). 
We assume that $t=s$ and $r+1 \leq s$. 
Then the inequality $st \leq 2t - (r+1) +s(s-1)$ holds.
It follows that 
\begin{align}\label{A:toprove(1)}
\frac{rst}{2} \leq rt - \frac{r(r+1)}{2} +r\frac{s(s-1)}{2}.
\end{align}
By (\ref{A:lowerboundCase1}), the right hand side of this inequality is a lower bound for the number of zeros in the row reduced form of the generator matrix (\ref{A:mainblock}). 
The left hand side of (\ref{A:toprove(1)}) is the half of the total number of entries in the row reduced form of the generator matrix (\ref{A:mainblock}). 
Therefore, we see that the generator matrix of $GHRS(\alpha,V,t-1)$ is a sparse matrix. 
Since the generator matrix of $GHRS(\alpha,V,t-1)$ is a parity-check matrix for the dual code $GHRS(\alpha,V,rs-t)$,
we conclude that $GHRS(\alpha,V,rs-t)$ is an LDPC code.

We proceed with the proof of (2). 
We assume that the inequality $st \geq  t^2 + t + r+1$ holds, which yields the inequality $s>t$ at once. 
It is also easy to check that the inequality $st \geq  t^2 + t + r+1$ is equivalent to the inequality 
\begin{align*}
rst- \frac{rt^2}{2} - \frac{rt}{2} - \frac{r(r+1)}{2} \geq \frac{rst}{2}.
\end{align*}
By (\ref{A:Case2}), the left hand side of our last inequality is exactly 
the number of zeros in the row reduced form of the generator matrix (\ref{A:mainblock}).
At the same time, the right hand side of the inequality is exactly the half of the total number of entries of the 
row reduced form of the generator matrix (\ref{A:mainblock}).
Therefore, we conclude as in the proof of (1) that 
the row reduced form of the generator matrix (\ref{A:mainblock}) is a sparse matrix. 
It follows that the corresponding parity check matrix of the dual code $GHRS(\alpha,V,rs-t)$ is sparse as well.
Hence, $GHRS(\alpha,V,rs-t)$ is an LDPC code.
This completes the proof of our theorem. 
\end{proof}

\begin{Example}

We fix our parameters: $q=17$, $s=7$, $r=3$, $t=3$.
Let $\alpha:=(3,2,7)$ be the list of evaluation points. 
Let $V$ the following multiplier matrix: 
\[
V:= 
\begin{bmatrix}
 8 &  9 & 10 \\
11 & 11 & 16 \\
11 &  2 & 11 \\
12 &  7 & 12 \\
 8 & 15 & 10 \\
 2 &  5 & 10 \\
10 &  4 & 16
\end{bmatrix}.
\]
After creating the codewords of $GHRS(\alpha,V,3)$ we convert them into row vectors in $\F_{17}^{21}$. 
It is easy to check that a generator matrix of the resulting code is given by 
\[
G:=
\begin{bmatrix}
\begin{array}{ccccccccccccccccccccc}
 1 &  0 &  0 & 13 & 12 &  0 &  9 & 14 &  9 &  0 &  0 &  0 &  0 &  0 &  0 &  0 &  0 &  0 &  0 &  0 &  0 \\
 0 &  1 &  0 & 14 &  6 &  2 &  4 & 10 &  4 &  0 &  0 &  0 &  0 &  0 &  0 &  0 &  0 &  0 &  0 &  0 &  0 \\
 0 &  0 &  1 &  4 &  3 &  7 &  9 & 14 &  9 &  0 &  0 &  0 &  0 &  0 &  0 &  0 &  0 &  0 &  0 &  0 &  0
  \end{array}
\end{bmatrix}.
\]
Then the corresponding parity-check matrix is given by 
\[
H:=
\begin{bmatrix}
\begin{array}{ccccccccccccccccccccc}
 1 &  0 &  7 &  0 &  0 &  4 &  0 &  0 & 15 &  0 &  0 &  0 &  0 &  0 &  0 &  0 &  0 &  0 &  0 &  0 &  0 \\
 0 &  1 & 12 &  0 &  0 &  8 &  0 &  0 &  0 &  0 &  0 &  0 &  0 &  0 &  0 &  0 &  0 &  0 &  0 &  0 &  0 \\
 0 &  0 &  0 &  1 &  0 & 11 &  0 &  0 &  8 &  0 &  0 &  0 &  0 &  0 &  0 &  0 &  0 &  0 &  0 &  0 &  0 \\
 0 &  0 &  0 &  0 &  1 & 11 &  0 &  0 & 10 &  0 &  0 &  0 &  0 &  0 &  0 &  0 &  0 &  0 &  0 &  0 &  0 \\
 0 &  0 &  0 &  0 &  0 &  0 &  1 &  0 & 16 &  0 &  0 &  0 &  0 &  0 &  0 &  0 &  0 &  0 &  0 &  0 &  0 \\
 0 &  0 &  0 &  0 &  0 &  0 &  0 &  1 &  6 &  0 &  0 &  0 &  0 &  0 &  0 &  0 &  0 &  0 &  0 &  0 &  0 \\
 0 &  0 &  0 &  0 &  0 &  0 &  0 &  0 &  0 &  1 &  0 &  0 &  0 &  0 &  0 &  0 &  0 &  0 &  0 &  0 &  0 \\
 0 &  0 &  0 &  0 &  0 &  0 &  0 &  0 &  0 &  0 &  1 &  0 &  0 &  0 &  0 &  0 &  0 &  0 &  0 &  0 &  0 \\
 0 &  0 &  0 &  0 &  0 &  0 &  0 &  0 &  0 &  0 &  0 &  1 &  0 &  0 &  0 &  0 &  0 &  0 &  0 &  0 &  0 \\
 0 &  0 &  0 &  0 &  0 &  0 &  0 &  0 &  0 &  0 &  0 &  0 &  1 &  0 &  0 &  0 &  0 &  0 &  0 &  0 &  0 \\
 0 &  0 &  0 &  0 &  0 &  0 &  0 &  0 &  0 &  0 &  0 &  0 &  0 &  1 &  0 &  0 &  0 &  0 &  0 &  0 &  0 \\
 0 &  0 &  0 &  0 &  0 &  0 &  0 &  0 &  0 &  0 &  0 &  0 &  0 &  0 &  1 &  0 &  0 &  0 &  0 &  0 &  0 \\
 0 &  0 &  0 &  0 &  0 &  0 &  0 &  0 &  0 &  0 &  0 &  0 &  0 &  0 &  0 &  1 &  0 &  0 &  0 &  0 &  0 \\
 0 &  0 &  0 &  0 &  0 &  0 &  0 &  0 &  0 &  0 &  0 &  0 &  0 &  0 &  0 &  0 &  1 &  0 &  0 &  0 &  0 \\
 0 &  0 &  0 &  0 &  0 &  0 &  0 &  0 &  0 &  0 &  0 &  0 &  0 &  0 &  0 &  0 &  0 &  1 &  0 &  0 &  0 \\
 0 &  0 &  0 &  0 &  0 &  0 &  0 &  0 &  0 &  0 &  0 &  0 &  0 &  0 &  0 &  0 &  0 &  0 &  1 &  0 &  0 \\
 0 &  0 &  0 &  0 &  0 &  0 &  0 &  0 &  0 &  0 &  0 &  0 &  0 &  0 &  0 &  0 &  0 &  0 &  0 &  1 &  0 \\
 0 &  0 &  0 &  0 &  0 &  0 &  0 &  0 &  0 &  0 &  0 &  0 &  0 &  0 &  0 &  0 &  0 &  0 &  0 &  0 &  1
 \end{array}
\end{bmatrix}.
\]

The Tanner graph defined by the parity check matrix $H$ is depicted in Fig.~\ref{fig:tanner-graph}.
\bigskip

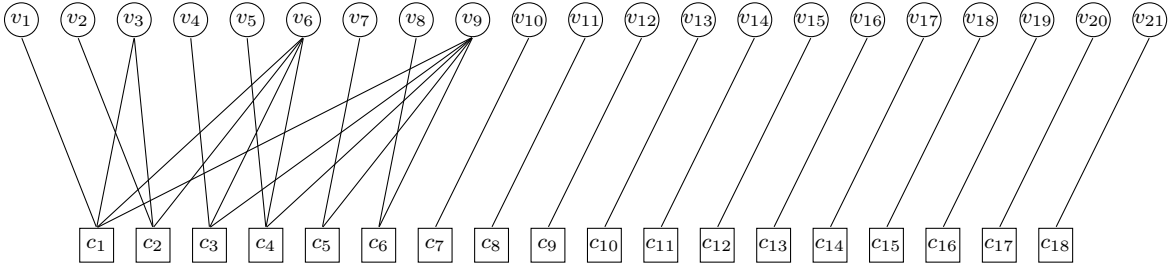
\begin{figure}[htbp]
    \centering

\begin{tikzpicture}[font=\scriptsize]

\tikzset{
  vnode/.style={circle, draw, minimum size=4.5mm, inner sep=0pt},
  cnode/.style={rectangle, draw, minimum size=4.5mm, inner sep=0pt},
  tedge/.style={-, line width=0.4pt, shorten >=0.4pt, shorten <=0.4pt}
}

\foreach \j in {1,...,21}{
  \node[vnode] (v\j) at (\j*0.75, 0) {}; 
  \node at (v\j) {$v_{\j}$};              
}

\foreach \i in {1,...,18}{
  \node[cnode] (c\i) at (1+\i*0.75, -3) {}; 
  \node at (c\i) {$c_{\i}$};              
}

\newcommand{\edgecv}[2]{%
  \draw[tedge] (v#2.south) -- (c#1.north);%
}

\edgecv{1}{1}
\edgecv{1}{3}
\edgecv{1}{6}
\edgecv{1}{9}

\edgecv{2}{2}
\edgecv{2}{3}
\edgecv{2}{6}

\edgecv{3}{4}
\edgecv{3}{6}
\edgecv{3}{9}

\edgecv{4}{5}
\edgecv{4}{6}
\edgecv{4}{9}

\edgecv{5}{7}
\edgecv{5}{9}

\edgecv{6}{8}
\edgecv{6}{9}

\foreach \i [evaluate=\i as \j using int(\i+3)] in {7,...,18}{
  \edgecv{\i}{\j}
}

\end{tikzpicture}
  \caption{Tanner graph corresponding to the parity-check matrix \(H\).}
    \label{fig:tanner-graph}
\end{figure}

Finally, we note that the sparsity of the generator matrix $G$ is 68.25\%, and the sparsity of the parity check matrix is 92.33\%

\end{Example}

\bigskip

To formalize the LDPC property of the GHRS codes established in  Theorem~\ref{intro:T3}, we now present a lemma that quantifies the sparsity of the associated parity-check matrices, followed by a refined proposition that consolidates these structural bounds.

\begin{Lemma}
\label{L:sparsity}
Let $C = GHRS(\alpha, V, t-1)$ be a GHRS code over $\mathbb{F}_q$, and let $H$ be a parity-check matrix for its dual code $GHRS(\alpha, V, rs - t)$. 
Then the following assertions hold true: 
\begin{enumerate}
    \item Each column of $H$ has weight at most $s$.
    \item Each row of $H$ has weight at most $cs$ for some constant $c$ independent of $r$.
\end{enumerate}
In particular, for fixed $s$ and sufficiently large $r$, the family of codes $GHRS(\alpha, V, rs - t)$ is LDPC.
\end{Lemma}

\begin{proof}
Each coordinate of a codeword in $GHRS(\alpha, V, rs - t)$ corresponds to an entry in an $s$-jet block, and hence to a column in the parity-check matrix $H$. Since each parity-check equation involves at most one $s$-jet per evaluation point, the number of nonzero entries per column is bounded above by $s$.

To bound the row weight, observe that each parity-check equation is derived from orthogonality with respect to the evaluation map $\text{Ev}_{\alpha,V}$, and thus involves a bounded number of jet blocks. From the structure of the generator matrix in (\ref{A:mainblock}), we know that each row of the generator matrix has support concentrated in a small number of jet blocks. Consequently, each parity-check equation (that is, each row of $H$) involves only a bounded number of jet blocks, each contributing at most $s$ nonzero entries. Therefore, the total number of nonzeros per row is bounded by $cs$ for some constant $c$ independent of $r$.
These observations show that both row and column weights of $H$ are bounded by constants depending only on $s$, and not on the blocklength $n = rs$. Hence, the code family is LDPC for fixed $s$ and growing $r$.
\end{proof}

\begin{Proposition}[Refined LDPC Property]\label{P:refinedLDPC}
Let $r, s, t \in \{l \in \mathbb{Z}: l \geq 2 \}$ with $t \leq rs - 1$, and let $\alpha = (\alpha_1, \dots, \alpha_r)$ be a list of $r$ distinct evaluation points in $\mathbb{F}_q$. Let $V = (v_{ij}) \in \text{Mat}_{s \times r}(\mathbb{F}_q)$ be a multiplier matrix with all entries nonzero. 
Then the dual code $GHRS(\alpha, V, rs - t)$ is LDPC under either of the following conditions:
\begin{enumerate}
    \item $t = s$ and $r + 1 \leq s$,
    \item $st \geq t^2 + t + r + 1$.
\end{enumerate}
Moreover, the associated parity-check matrix $H$ satisfies the sparsity bounds given in Lemma~\ref{L:sparsity}, ensuring that both row and column weights remain bounded as $r \to \infty$.
\end{Proposition}

\begin{proof} 
We will show that the parity-check matrix $H$ of the dual code $GHRS(\alpha, V, rs - t)$ has bounded row and column weights, thereby satisfying the LDPC criteria.

\emph{Column weight bound:} Each coordinate of a codeword corresponds to an entry in an $s$-jet block. Since each parity-check equation involves at most one $s$-jet per evaluation point, the number of nonzero entries per column is at most $s$. This bound is independent of the blocklength $n = rs$.

\emph{Row weight bound:} Each parity-check equation corresponds to a linear constraint orthogonal to the row space of the generator matrix of $GHRS(\alpha, V, t-1)$. From the block structure of the generator matrix (see (\ref{A:mainblock})), each row of the generator matrix has support concentrated in a small number of jet blocks. Consequently, each parity-check equation involves only a bounded number of jet blocks, each contributing at most $s$ nonzero entries. Therefore, the total number of nonzeros per row is bounded by $cs$ for some constant $c$ independent of $r$.

\emph{Asymptotic LDPC behavior:} For fixed $s$, as $r \to \infty$, the blocklength $n = rs$ grows, but the row and column weights of $H$ remain bounded. This shows that the family of codes $GHRS(\alpha, V, rs - t)$ is LDPC for fixed $s$ and sufficiently large $r$.
\end{proof}

To further illustrate the LDPC conditions from Theorem~\ref{intro:T3} (or Proposition~\ref{P:refinedLDPC}), we present a table showing the minimum values of $r$ for various choices of $s$ and $t$.

\begin{table}[h!]
\centering
\caption{Minimum values of $r$ for which GHRS codes satisfy LDPC conditions from Theorem~\ref{intro:T3}}
\begin{tabular}{|c|c|c|c|}
\hline
$s$ & $t$ & Min $r$ (Cond 1: $t = s$, $r \leq s - 1$) & Min $r$ (Cond 2: $r \leq st - t^2 - t - 1$) \\
\hline
2 & 2 & 1 & -- \\
3 & 3 & 2 & -- \\
4 & 4 & 3 & -- \\
5 & 5 & 4 & -- \\
6 & 3 & -- & 4 \\
6 & 4 & -- & 9 \\
6 & 5 & -- & 14 \\
6 & 6 & 5 & 19 \\
7 & 3 & -- & 6 \\
7 & 4 & -- & 13 \\
7 & 5 & -- & 20 \\
8 & 4 & -- & 15 \\
8 & 5 & -- & 24 \\
9 & 5 & -- & 28 \\
10 & 6 & -- & 37 \\
10 & 10 & 9 & -- \\
\hline
\end{tabular}
\label{tab:ldpc_thresholds}
\end{table}

\section{Quasi-cyclic GHRS Codes}
\label{S:Quasicyclic}

Let $\alpha\in\F_q^\times$ and suppose $u=(1,\alpha,\alpha^2,\dots,\alpha^{r-1})$ is a geometric progression.
Then the code $GHRS(u,V,t-1)$ is quasi-cyclic of index $r$: a cyclic shift of the $r$ block-columns maps codewords to codewords.
Equivalently, there exists a polynomial parity-check matrix $H(D)$ with $r\times r$ blocks over $\F_q[D]$ such that $\mathcal{C}=\{x(D)\in (\F_q[D]/(D^r))^{s}\mid H(D)\,x(D)^\top=0\}$.
This is a direct consequence of the multiplicative relation $\partial^{i-1}g(\alpha^{j})=\alpha^{(i-1)}\,\partial^{i-1}g(\alpha^{j-1})$,
where $g(x) = f(x/\alpha)$, recorded in the proof of Theorem~\ref{intro:T4}, which intertwines with the block-cyclic permutation operator.

Every (linear) code in $\Mat_{r \times s}(\F_q)$ can be naturally viewed as a (linear) code in $\F_q^{rs}$
by converting matrices into row vectors, reading the entries column-by-column from top to bottom, starting from the first column and moving to the next. 
To distinguish between these two interpretations of the same code, if $\mc{C}$ is a subset of $\Mat_{r \times s}(\F_q)$, we denote the corresponding subset of $\F_q^{rs}$ by $\widehat{\mc{C}}$.
In this vein, for an element $A\in \mc{C}$, the corresponding row vector in $\widehat{\mc{C}}$ will be denoted by $\hat{A}$. 

\begin{Example}
Let $A$ be the following matrix:
$$
A:=\begin{bmatrix}
a_{11} & a_{12} & a_{13} \\
a_{21} & a_{22} & a_{23} \\
a_{31} & a_{32} & a_{33} \\
\end{bmatrix}.
$$
Then the corresponding row vector is given by 
$$
\hat{A}:= (a_{11} , a_{21} , a_{31} , a_{12} , a_{22} , a_{32} , a_{13} , a_{23} , a_{33} ).
$$
\end{Example}

\medskip

The following notation for representing matrices will be useful for our purposes.
If the columns of the matrix $A$ are given by $A_1,\dots, A_r$, then we write 
$$
[{\rm Col}_1(A), \ldots , {\rm Col}_r(A)]
$$ 
to denote $A$. 
We will call a (linear) code $\mc{C}$ in $\Mat_{r\times s}(\F_q)$ a \emph{column-cyclic code} if for any matrix $A$  
in $\mc{C}$, the matrix obtained by cyclic shift of its columns, 
\[
T^1_{\Mat_{r\times s}(\F_q)}(A) := [{\rm Col}_r(A),{\rm Col}_1(A), \ldots , {\rm Col}_{r-1}(A)], 
\]
is also in $\mc{C}$. 
\medskip

\begin{Lemma}\label{L:columncyclic}
Let $\mc{C}$ be a code in $\Mat_{r\times s}(\F_q)$. 
Then $\mc{C}$ is a column-cyclic code if and only if the corresponding code $\widehat{\mc{C}}$ in $\F_q^{rs}$ is a quasi-cyclic code of index $r$. 
\end{Lemma}

\begin{proof}
The proof of this lemma follows from the definitions and will be skipped. 
\end{proof}

\medskip

In~\cite{GuneriLingOzkaya}, quasi-cyclic (QC) codes of index $r$ are defined by organizing codewords in $\F_q^{rs}$ into $r \times s$ matrices, where each matrix is formed by reading the codeword entries in $r$ consecutive blocks and placing them as rows. A QC code is characterized by the closure of these matrices under cyclic row shifts.
When using the Hamming weight, the code $\mc{C}$ and its transpose $\mc{C}^\top$ are essentially equivalent because the transposition map $A \mapsto A^\top$ is a linear isometry between $\Mat_{r \times s}(\F_q)$ and $\Mat_{s \times r}(\F_q)$. Thus, row or column shift invariance does not matter.
However, with the NRT metric, the transposition map is not an isometry. Therefore, matrices are translated into row vectors by reading them column by column, from top to bottom and left to right, to maintain the correct metric properties.
\medskip

We are now ready to prove our theorem on quasi-cyclicity of the Generalized Hyperderivative Reed-Solomon codes. 
Let us recall the statement of our Theorem~\ref{intro:T4} from the introductory section. 
\medskip

Let $\alpha\in\mathbb{F}_q^*$ be such that $\{1, \alpha, \alpha^2, \dots, \alpha^{r-1}\}$ is a cyclic subgroup of $\F_q^*$. 
Let $u:=(1, \alpha, \alpha^2, \dots, \alpha^{r-1})$.
Let $V:=(v_{ij})_{i=1,\dots, s\atop j=1,\dots, r}$ be an $s\times r$ multiplier matrix such that 
\[
\frac{v_{ij}}{v_{i,j-1}} = \alpha^{\, i-1} \quad \text{ for all $i=1,\dots,s$ and $j=1,\dots,r$},
\]
with the convention that $v_{i,0} = v_{i,r}$. 
Then $GHRS(u,V,t-1)$ is a quasi-cyclic code of index $r$. 
\medskip

\begin{proof}[Proof of Theorem~\ref{intro:T4}]
To ease our notation, let us denote by $\mathcal{D}$ the Generalized Hyperderivative Reed-Solomon code $GHRS(u,V,t-1)$,
where $u$ and $V$ are as in the hypotheses of the theorem. 
Then a matrix 
    \[
    A = \bigl(v_{ij}\, a_{ij}\bigr)_{i=1,\dots, s\atop j=1,\dots, r}
    \]
    is a codeword in the code $\mathcal{D}$ if there exists a polynomial $f(x)$ of degree at most $t-1$ such that
    \[
    \partial^{i-1} f(\alpha^{j-1}) =  a_{ij} 
    \]
    for every $i=1,\dots, s$ and $j=1,\dots, r$. 
We will show that if we replace ${\rm Col}_j(A)$ by ${\rm Col}_{j+1 \mod r}(A)$, the resulting matrix is still in $\mathcal{D}$.
To this end, let $A'$ denote the matrix obtained from $A$ by shifting each column of $A$ one step to the right (with the last column wrapping to the first):
\[
A' = \Bigl( v_{ij}\, a'_{ij}\Bigr)_{i=1,\dots, s\atop j=1,\dots, r}
\]
where
\[
a'_{ij} = a_{i,j-1} \quad \text{with the convention } a_{i,0}=a_{i,r}.
\]
Our goal is to show that there exists a polynomial $g(x)$ of degree at most $t$ such that
\[
\partial^{i-1} g(\alpha^{j-1}) =  a'_{ij} =  a_{i,j-1} \quad \text{for all }\ {1\le i\le s,\,1\le j\le r}.
\]
We begin with defining 
\[
g(x) :=  f\Bigl(\frac{x}{\alpha}\Bigr).
\]
Note that, since $\alpha\neq 0$, $g(x)$ is also a polynomial of degree at most $t-1$.

Using the chain rule for hyperderivatives we obtain
\[
\partial^{i-1}g(x) = \alpha^{-(i-1)}\, \partial^{i-1}f\Bigl(\frac{x}{\alpha}\Bigr).
\]
Evaluating at $x=\alpha^{j-1}$ gives
\[
\partial^{i-1}g(\alpha^{j-1}) = \alpha^{-(i-1)}\, \partial^{i-1}f\Bigl(\frac{\alpha^{j-1}}{\alpha}\Bigr)
= \alpha^{-(i-1)}\, \partial^{i-1}f(\alpha^{j-2}).
\]
That is,
\[
\partial^{i-1}g(\alpha^{j-1}) = \alpha^{-(i-1)}\, a_{i,j-1}.
\]

Now, multiplying by the multiplier $v_{ij}$, the weighted entry in the shifted matrix becomes
\[
v_{ij}\, \partial^{i-1}g(\alpha^{j-1}) = v_{ij}\, \alpha^{-(i-1)}\, a_{i,j-1}.
\]
By our assumption on the multipliers, we have
\[
\frac{v_{ij}}{v_{i,j-1}} = \alpha^{\, i-1} \quad \Longrightarrow \quad v_{ij}\,\alpha^{-(i-1)} = v_{i,j-1}.
\]
Thus,
\[
v_{ij}\, \partial^{i-1}g(\alpha^{j-1}) = v_{i,j-1}\, a_{i,j-1}.
\]
But the right-hand side is exactly the weighted entry in column $j-1$ of the original codeword. Hence, the cyclically shifted matrix
$A'$ is given by
\[
A' = \bigl(v_{ij}\, \partial^{i-1}g(\alpha^{j-1})\bigr)_{i=1,\dots, s\atop j=1,\dots, r}.
\]
This shows that $A'$ is a codeword in $\mathcal{D}$.
Hence the code $\mathcal{D}$ is invariant under cyclic shifts of its columns.
This finishes the proof of our assertion.
\end{proof}

\section{Closing Remarks and Questions}\label{S:Appendix}

As we mentioned in the introduction, (function field) analogs of the Reed-Solomon codes in the NRT metrics are extensively investigated by many authors.
In particular, Niederreiter and Xing~\cite{NX} considered them in the context of digital nets. 
In~\cite{NiOz2002}, Niederreiter and \"Ozbudak developed a far reaching generalization of the results of~\cite{NX},
where they used arbitrary places not just rational places. 
It would be mathematically very interesting to extend our results to the algebro-geometric setting of the article~\cite{NiOz2002}.  
More recently, Can, Montero, and \"Ozbudak defined introduced analogs of the Reed-Solomon codes by using certain subspaces of $\Mat_{s\times r}(\F_q)$ and certain metrics (called the bottleneck metrics) that are closely related to the NRT metrics. It would also be very interesting to investigate to what extent the results of the current article adopts to the bottleneck metric Reed-Solomon codes. 
\medskip

In~\cite{Jensen}, Jensen showed that a quasi-cyclic code can be written as a direct sum of concatenated codes, where the inner codes are minimal cyclic codes and the outer codes are linear codes.
Advancing Jensen's work, G\"uneri and \"Ozbudak~\cite{GO2013} showed that the outer codes are nothing but the constituents of the quasi-cyclic code in the sense of the work~\cite{LingSole} of Ling and Sol\'e.
It would be very interesting to determine Jensen decomposition of the GHRS codes in the sense of Ling and Sol\'e. 
\medskip

There is a fascinating interplay between NRT-codes and ordered orthogonal arrays.
This connection was first investigated by Barg and Purkayastha~\cite{BargPurkayastha}.
It would be very interesting to understand the ordered orthogonal arrays corresponding to Hyperderivative Reed-Solomon codes. 
\medskip

In~\cite{BargPark}, Barg and Park investigated the multivariate Tutte polynomials, higher Hamming weights, as well as poset matroids of the NRT-It would be interesting to calculate the multivariate Tutte polynomials of the GHRS codes following the work of Barg and Park. 
\medskip

Under different conventions and notation, the automorphism group of the NRT metric on $\Mat_{s\times r}(\F_q)$ is determined by Lee in~\cite{Lee2003}. Let $\mathbb{B}_s^-$ denote the Borel group of all $s \times s$ lower triangular invertible matrices with entries from $\F_q$. For a matrix $A \in \Mat_{s\times r}(\F_q)$ and $j \in \{1, \dots, r\}$, let ${\rm Col}_j(A)$ denote the $j$-th column of $A$. The action of the product group $(\mathbb{B}_s^-)^r$ on $\Mat_{s\times r}(\F_q)$ is defined by 
\begin{align}\label{A:B^-action}
(B_1, \dots, B_r) \cdot A = [B_1 \cdot {\rm Col}_1(A) : \ldots : B_r \cdot {\rm Col}_r(A)],
\end{align}
where $(B_1, \dots, B_r) \in (\mathbb{B}_s^-)^r$.
Here, $B_j \cdot {\rm Col}_j (A)$, $1\leq j \leq r$, is the usual matrix multiplication action of $\mathbb{B}_s^-$ on the column vectors.
Additionally, there is a natural action of the symmetric group $S_r$ on $\Mat_{s\times r}(\F_q)$ which is given by  
\begin{align}\label{A:Saction}
\sigma \cdot A = [{\rm Col}_{\sigma(1)}(A) : {\rm Col}_{\sigma(2)}(A) : \ldots : {\rm Col}_{\sigma(r)}(A)],
\end{align}
where $\sigma \in S_r$. 
Evidently, the two actions (\ref{A:B^-action}) and (\ref{A:Saction}) commute with each other,
implying that wreath product $(\mathbb{B}_s^-)^r\wr S_r$ acts on matrices: 
\begin{align*} 
(\mathbb{B}_s^-)^r\wr S_r \ \ \times \ \  \Mat_{s\times r}(\F_q) & \longrightarrow \Mat_{s\times r}(\F_q) \\
(((B_1, \dots, B_r), \sigma ) \ ,\  A  )\hspace{.75cm} & \longmapsto [B_1\cdot {\rm Col}_{\sigma^{-1}(1)}(A) :  \ldots : B_r\cdot {\rm Col}_{\sigma^{-1}(r)}(A)].
\end{align*}
It is easy to check that this action preserves the NRT metric. 
In fact, as we mentioned earlier, Lee shows in~\cite{Lee2003} that the group of linear isometries of $\Mat_{s\times r}(\F_q)$ 
with respect to the NRT metric is $(\mathbb{B}_s^-)^r\wr S_r$.
We note also that the wreath product $(\mathbb{B}_s^-)^r\wr S_r$ is isomorphic to the semidirect product $(\mathbb{B}_s)^r\rtimes S_r$,
where $\mathbb{B}_s$ is the group of all $s\times s$ invertible upper triangular matrices with entries from $\F_q$.
\medskip

It is shown in the references~\cite{Skriganov2007} and~\cite{HyunLee2011} that, for a given linear MDS poset code $\mc{C}$ over $\F_q$, the orbit of $\mc{C}$ under the action of the full linear isometry group contains codes that meet the Gilbert-Varshamov bound for their Hamming weights. 
In this regard, it would be of significant interest to determine the stabilizer subgroups in ${\rm Aut}(\Mat_{s\times r}(\F_q))$ of all GHRS codes.

\section*{Acknowledgements}

The authors thank the Louisiana Board of Regents for their support through the grant LEQSF(2023-25)-RD-A-21.
The authors also thank the referees for their careful reading and constructive comments which significantly improved the quality of the paper.

\bibliographystyle{plain}
\bibliography{references}

\end{document}